\documentclass[12pt]{article}
\usepackage{graphics}
\input epsf
\usepackage{graphics}
\usepackage[dvips]{color}
\usepackage{multicol}

\usepackage{amsmath,color,fancybox,graphics,graphpap,rotating}
\definecolor{white}{rgb}{1,1,1}
\definecolor{yellow}{rgb}{0.95,0.75,0.1}
\definecolor{red}{rgb}{0.5,0,0}
\definecolor{green}{rgb}{0,1,0}
\definecolor{blue}{rgb}{0,0.5,1}
\definecolor{bgcolor}{rgb}{0.94,0.91,0.78}

\definecolor{lblue}{rgb}{0,0.8,1}
\definecolor{dblue}{rgb}{0,0,.6}
\definecolor{dgreen}{rgb}{0,0.3,0}
\definecolor{lila}{rgb}{0.8,0,0.8}
\definecolor{violet}{rgb}{1,0,1}
\definecolor{grey}{rgb}{0.3,0.3,0.3}
\definecolor{turquoise}{rgb}{0,.9608,1}

\definecolor{contoura}{rgb}{0,0,1}
\definecolor{contourb}{rgb}{0,1,1}
\definecolor{contourc}{rgb}{0,1,0}
\definecolor{contourd}{rgb}{0.95,0.75,0.1}
\definecolor{contoure}{rgb}{1,0,0}
\definecolor{contourf}{rgb}{1,0,1}

\def\lsim{\raise0.3ex\hbox{$\;<$\kern-0.75em\raise-1.1ex\hbox{$\sim\;$}}}
\def\gsim{\raise0.3ex\hbox{$\;>$\kern-0.75em\raise-1.1ex\hbox{$\sim\;$}}}

\newcommand{\nee}{\nonumber \end{eqnarray}}

\newcommand{\be}{\begin{eqnarray}}
\newcommand{\ben}{\begin{eqnarray}\nonumber}
\newcommand{\ee}{\end{eqnarray}}

\begin{document}

\title{An Alternative String Landscape Cosmology:\\ Eliminating Bizarreness}
\author{
L. Clavelli\footnote{lclavell@bama.ua.edu}$\;$\footnote{on leave from
Dept. of Physics and Astronomy, Univ. of Alabama, Tuscaloosa AL 35487} and
Gary R. Goldstein\footnote{gary.goldstein@tufts.edu}\\
Dept. of Physics and Astronomy, Tufts University, Medford MA 02155}

\date{}
\maketitle

\begin{abstract}
In what has become a standard eternal inflation picture of the string landscape there are many problematic consequences and a difficulty defining
probabilities for the occurrence of each type of universe.  One feature in particular that might be philosophically disconcerting is the infinite cloning of each individual and each civilization in infinite numbers of separated regions of the multiverse.  Even if this is not ruled out due to causal separation one should ask whether the infinite cloning is a universal prediction of string landscape models or whether there are scenarios in which it is avoided.  If a viable alternative cosmology can be constructed one might search for predictions that might allow one to discriminate experimentally between the models.  We present one such scenario although, in doing so, we are forced to give up several popular presuppositions including the absence of a preferred frame and the homogeneity of matter in the universe.
The model also has several ancillary advantages.  We also consider the future lifetime of the current universe before becoming a light trapping region.  
\end{abstract}
keywords: String Landscape, String cosmology, Multiverse, Supersymmetry, preferred frame
\section{Introduction}
     By now it is widely known that Einstein's greatest blunder was not, as he thought, the introduction of the cosmological constant but rather his uncritical presupposition that the universe was static and eternal into the past and into the future.  It might be that the eternal inflation (or chaotic inflation) model \cite{Linde07} rests on a similar untested presupposition that the universe has neither beginning nor end and is
an isotropic, chaotically bubbling but otherwise homogeneous vat of varying vacuum energies and matter content.
  
In the standard picture it is difficult to define probabilities \cite{GuthVanchurin}
and difficult to prevent the occurrence of infinite cloning of each individual as well as the production of disembodied brains (Boltzmann brains \cite{BoltzmannBrains}) and  other bizarre entities. { In a homogeneous and isotropic multiverse, everything that can exist does exist in infinite numbers.}
There is also an entropy problem in which, if the universe had an infinite past, one might wonder why our universe is not totally featureless. 
We ask whether one can construct a viable landscape scenario that avoids the paradoxes and bizarre features of the eternal inflation scenario.
For example, can one avoid an infinite number of human, quasi-human, and monster populations in the string landscape and can one avoid the other problems mentioned above?     

If, as is commonly assumed, the discovery of dark energy is due to our existence in a De Sitter space-time,  the universe is homogeneous and isotropic with a scale factor
growing at large times as an exponential of t.  In this world the matter density is homogeneous and varies as the inverse cube of the scale factor $a(t)$.  

\be
    \lim_{t\rightarrow \infty} a(t) = e^{H t}
\ee
\be
    \rho_m(t) \approx {a(t)}^{-3} \quad .
\label{FRW}
\ee

A homogeneous De Sitter space, of course, is not compatible with the notion that each universe within the multiverse is a spherical bubble with a center.  We seek to construct a bubble universe
that retains, in an adequate approximation, the results of eq.\,\ref{FRW}. {This necessarily
involves a preferred frame, the bubble center, and implies some amount of matter inhomogeneity.}

If our model, like the standard cosmology, is not in conflict with observations, each physicist is entitled to assume as a presupposition either model based on perceived aesthetic differences keeping in mind that the alternative is not ruled out. In addition to being consistent with current observations we point out several phenomenological advantages of our picture including the banishing of infinite numbers of monsters from the multiverse.  In this we may be pursuing a time-honored role of science.

Next we briefly enumerate nine assumptions.  The number of assumptions may seem too great to be justified by the mere avoidance of infinite cloning and other landscape paradoxes.  However, if these assumptions are not ruled out and can be distinguished in their consequences from those of the standard inflation picture, we feel that they should be critically considered.  

In a subsequent section   
we elaborate on the assumptions and examine their consequences relative to the more standard string landscape picture.  Finally, in a concluding section, we summarize the advantages of our proposed model. 

{\bf We assume:}
\begin{enumerate}
\item
{\bf The multiverse is not infinite into the past but originated at some time $t=-t_0$.}  
\item{\bf There is a finite rest energy of matter $M c^2$ in the multiverse centered on an origin.} 
\item{\bf At time $t=-t_0$ the matter density is proportional to a delta function at the spatial origin.}
\item{\bf The multiverse begins at $t=-t_0$ in a state of high vacuum energy density.}  It then rapidly cascades down to the present state of vacuum energy
\be
     \rho_v = 4.0 \pm 0.2 \; \displaystyle{GeV/c^2 /m^3} \quad .
\ee
\item{\bf These transitions are governed by the Coleman-De Luccia thin wall formulae and transitions up to higher vacuum energy are of negligible probability.}

\item{\bf Since such a phase transition does not create particles only a transition to our $\rho_v$ near the peak of the matter distribution ($r=0$) will lead to galaxies, planets, and people.}
\item{\bf We are protected from being enveloped in a black hole like region by a future phase transition to zero vacuum energy.}  
\item{\bf Eventually, a bubble of zero vacuum energy should form within our universe and expand to convert the entire universe to this presumably supersymmetric state.}  
\item{\bf This zero vacuum energy state is the ground state of the multiverse.}
\end{enumerate} 

\section{Discussion}
We now reiterate the assumptions and explain their relevance and effect to compare with the standard eternal inflation picture.

\begin{enumerate}
\item
{\bf The multiverse is not infinite into the past but originated at some time $t=-t_0$.}  There is a growing consensus that the alternative leads to numerous thorny paradoxes \cite{Vilenkin}. The maintainance of a
universe from the infinite past requires that there is an equilibrium between transitions to lower vacuum energy and transitions to higher vacuum energy.  However, in a thermodynamic system, transitions to lower energy must be enormously favored over {the entropy decreasing} transitions to higher states {and asymptotically in time the system should preferentially settle into the lowest available state.}  
\item{\bf There is a finite rest energy of matter $M c^2$ in the multiverse centered on an origin.}  This assumption may be initially repugnant to most cosmologists but is not ruled out.  
It suggests, contrary to what physics has held for a century, that there is a preferred frame in the multiverse. The assumption suggests that
inertial frames are those that are unaccelerated relative to this origin.  
This allows rotating coordinating systems to be inequivalent.  In classical physics and in general relativity it is taught that the laws of physics hold in inertial frames but no prescription is given a priori for {defining} an inertial frame.  The best that can be done is to state that, if a frame is found in which Newton's laws hold, they will also hold in any frame traveling relative to that frame with constant velocity.  Newton, himself, was aware of the puzzle and posed the famous ``bucket problem". A rotating bucket of water has a concave surface even when viewed from a co-rotating frame. The current assumption is analogous to Mach's Principle which states that the matter distribution of the universe defines a preferred frame, in his case the rest frame of the distant stars.  Given a finite total mass, there can be no infinite cloning of individuals.
Similarly, the Boltzmann Brain problem of standard string cosmology is largely due to the infinite number of universes like ours.
{Thermal fluctuations are a property of material substances.}  
Given that only a vanishing fraction of the multiverse contains appreciable matter in our model, thermal fluctuations are unlikely to create Boltzmann Brains and quantum fluctuations are unlikely to produce such macroscopic complex objects in the short time available.  
{  
Given the normal laws of evolution, it is easier to create a complete human being than a disconnected brain and this preference may become absolute if there is only a limited amount of matter available.}
The number of Boltzmann Brains with macroscopic baryon number produced in our model in matter-containing regions due to quantum fluctuations in the short time since the big bang should not be expected to be large compared to the vanishing number we have already observed.
\item{\bf In the limit $t \rightarrow -t_0$ the matter density is proportional to a delta function at the spatial origin.}
With two free parameters, $M$ and $R$, we may write {as a non-unique example}
\be 
    \rho_m (r,t) = \frac{M}{(R {\sqrt\pi}\,a(t))^3} e^{-r^2/(Ra(t))^2} \quad .
\label{rhom}
\ee
The scale factor $a(t)$ is an increasing function of time that vanishes at $t=-t_0$ and that can be taken equal to unity at the current time, $t=0$.  The equation of continuity
\be
     \vec{\nabla} \cdot (\rho_m \vec{v}(r)) + \frac{\partial \rho_m }{\partial t} = 0
\label{eqofcontinuity}
\ee
has, independent of $R$, the solution
\be
     \vec{v}(r) = \vec{r}\; \frac{\dot{a}}{a}  
\label{HubbleFlow}
\ee
in agreement with Hubble's law.  This justifies the identification of $a(t)$ with the scale factor.
An important open question left to future study is the effect of curvature corrections to 
these equations and other results of this work.  {In classical and quantum physics the initial conditions do not follow from physical law although they must be among the possible states of the system.  Thus one cannot require a justification for initial conditions such as we propose in the preceeding assumptions.  In standard cosmology there is no comparably clear statement of initial conditions.}

As $t$ approaches $-t_0$ the matter density of \ref{rhom} approaches a spatial delta function at the origin 
\be
     \rho_m(\vec{r},-t_0) &=& M \delta^3(\vec{r})
\label{delta}
\ee
and it integrates to a time independent mass
\be
     \int d^3 r \,\rho_m(r,t) = M  \quad .
\label{Mtot}
\ee

$\rho_m$
can be taken to be a neutral, flavor and color singlet density of quarks, leptons, and gauge bosons together with their broken susy partners.  
Apart from pair production at positive times, there is no anti-matter in the universe.  
This possible solution to the baryon asymmetry problem is not available in the standard picture where matter is pair produced at the end of the inflationary era.  {Of course, one is free to consider, in our model also, the possibility that the matter density at the origin of time is a CP symmetric state, thus
preserving a possible role for CP violation at later times and the interesting problem of insuring an asymmetric survival of a sufficient number of baryons.}

The matter distribution is isotropic relative to its origin.  At $r/R  << a(t)$ and at $r/R >> a(t)$ the matter density is also homogeneous so the metric approaches a Friedmann, Robertson, Walker (FRW) metric and the matter density satisfies eq.\,\ref{FRW}.  At sufficiently large $t$, the density is homogeneous (and for large $r$ negligible) around all fixed positions.  A prediction of this picture is that significant deviations from  homogeneity should exist near $r/R \approx a(t)$.  
{Observations from Earth cannot probe inhomogeneities at scales larger than the Hubble length, the distance light has travelled since the big bang.  Thus if
$R$, is much greater than the Hubble length, the proposed matter distribution is indistinguishable from the homogeneous distribution of the standard cosmological model.  This will be our default assumption although, in future work, one might ask whether smaller values of $R$ might also be observationally viable.  Of course if we choose $R$ to be beyond current limits of observability, the attractiveness of the model, though not its viability, must depend on other conceptual advantages.}

{The cosmic background radiation is isotropic in its rest frame.  Relative to this frame the earth frame is presently moving with a speed of
$369\,\pm 1\,$km/s.  At present the origins of these two frames are $5.1$ Mpc apart.  
One way to
subtract the dipole term from the CBR is to cite all measurements as they would appear in the CBR rest frame.  If the CBR rest frame is the $r=0$ position of our model the
universe is isotropic relative to this frame but not homogeneous except on scales small compared to our $R$ parameter.} 

{On the other hand if the $r=0$ position is far from the center of our sphere of last scattering, the universe would still appear isotropic and homogeneous if $R$ is much greater than the Hubble length, $4.2$ Gpc.} 
A direct measure of how homogeneous the universe is on large scales is a complicated issue
\cite{Papai}.  

If the sun is close to the $r=0$ position, it could provide a basis for understanding the remarkable large angle correlations \cite{Schwarz} between the ecliptic plane and the CBR.  (The great statistical significance of these correlations, however, could be somewhat weakened by systematic effects \cite{InoueSilk}).
Then, if $R$ is in the Gpc range one would have to wonder whether some selection effect led to the
closeness of the solar system to the origin or, equivalently, to the smallness of the solar velocity relative to the CBR.  E.g. perhaps the larger collision rate of fast moving inhomogeneities with the matter background inhibits the rise of life.  Alternatively it could be related to galactic merger rates as a function of velocity.  These could be residual effects related to the anthropic discrimination against high inflation rates \cite{Weinberg}. 

Given a small velocity,  
the parameters, $M$ and $R$, in eq.\,\ref{rhom} are constrained by the observed current matter density and degree of matter homogeneity.  
If at the current time the averaged matter distribution is homogeneous out to a Hubble length and falls off rapidly thereafter, 
the current value of the matter density, $\rho_m(0) = \Omega_m(0)\,\rho_c = 1.4\pm0.2 \,\displaystyle{GeV/c^2/m^3}$ then implies that the rest energy of the universe is
\be
      M\,c^2 \approx \rho_m(0)\,c^2\,\pi^{3/2} R^3 > 2.6 \cdot\,10^{76} \,\displaystyle{GeV} \quad .
\ee


Newton's laws hold to an excellent approximation implying, in the current context, that the earth and the milky way are presently moving with negligible acceleration and with some velocity, $\vec{v}$ relative to the origin.   The magnitude of $\vec{v}$ could be related to the dipole term that is evident in the cosmic background radiation (CBR) or could be some other speed.
The matter distribution relative to earth is then
\be
    \rho_m (\vec{r},t) = \frac{M}{(R {\sqrt\pi}\,a(t))^3} e^{-(\vec{r}+\vec{v}(t+t_0))^2 /(Ra(t))^2} \quad .
\label{dipole}
\ee
Here $\vec{v}$ is not to be confused with the Hubble flow of eq.\,\ref{HubbleFlow}.
Small deviations from isotropy beyond the dipole term can be obtained by multiplying this equation or that of eq.\,\ref{rhom} by the spherical harmonic expansion
\be
     F(r,t) = 1 + \sum_{l,m} f_{l,m}(r,t) Y_{l,m}(\theta,\phi) \quad .
\ee
The acoustic peak analysis suggests that the lowest values of $l$ for which $f_{l,m}$ are
appreciable are near $l\approx 100$.  Expanding eq.\,\ref{dipole} in the velocity $\vec{v}$
produces correlated low multipole moments suppressed by inverse powers of $R$.  The quadrupole moment, for instance, is suppressed in agreement with observation whereas this suppression is problematic in the standard cosmology.

If the $f_{l,m}$ vanish at $r=0$ and are sufficiently well behaved in $t$ and at large $r$, the properties of eqs.\,\ref{delta} and \ref{Mtot} are preserved.

\item{\bf The multiverse begins at $t=-t_0$ in a state of high vacuum energy density, $(M_I c^2)^4/(\hbar c)^3$.}  It then rapidly cascades down to the present state of vacuum energy
\be
     \rho_v = 4.0 \pm 0.2 \, \displaystyle{GeV/c^2 /m^3} \quad .
\ee

\item{\bf These transitions are governed by the Coleman-De Luccia \cite{CdL} thin wall formulae and transitions up to higher vacuum energy are of negligible probability.}
In the thin wall approximation, the energy released in the transition to a lower vacuum energy does not create particles but goes into the surface energy.
In the words of Coleman \cite{Coleman} ``This refutes the naive expectation that the decay of the false vacuum would leave behind it a roiling sea of mesons".
The thin wall assumption thus implies that galaxies, planets, and people come not from vacuum decay but from a primordial matter distribution such as that of eq.\,\ref{rhom}.

The probability per unit time per unit volume to nucleate a bubble of critical size destined to take over the universe is
\be
     \frac{dP}{dt\,d^3r} = A e^{-B}
\label{CdL}
\ee
with
\be
         B  = \frac{27 \pi^2 S^4}{2 \hbar c(c^2 \Delta \rho)^3}
\label{B}
\ee
where $S$ is the energy per unit area on the surface.

The two parameters $A$ and $S$ must at this point be considered free since they are determined by the exact shape of the unknown effective potential.  One can speculate about various values for these parameters.  For example, if supernovae Ia are triggered by a transition \cite{BC} to exact supersymmetry (zero vacuum energy) within a
white dwarf one arrives at the estimates
\ben
     S &\approx& 4.76\cdot 10^{34}\, \displaystyle{GeV/m^2}\\
     A^{-1} &\approx&  7.89\cdot 10^{36}\,\displaystyle{m^3 s} \quad .
\label{estimates}
\ee

However, it has been suggested \cite{Frampton} that $A^{-1}$ depends on $\Delta \rho$.
The wall thickness in the transition out of inflation can be estimated by equating the energy released in a shell of thickness $\delta r$ to the increase of energy stored in the wall.
\be
      4 \pi r^2 \Delta\rho\, c^2 \delta r = 4 \pi r^2 S \quad .
\ee
If $\Delta\rho$ is anywhere near the GUT density ${M_G}^4 c^3 /\hbar^3$ and $S$ is near the estimate of eq.\,\ref{estimates}, the wall thickness is extremely small.
\be
      \delta r \approx 3.64 \cdot 10^{-77} \,\displaystyle{m} \quad .
\ee
At the time of bubble nucleation this is $1/3$ of the bubble radius and becomes rapidly negligible in comparison as the bubble grows.

Also, in the case of the estimates of \cite{BC},
$B$ is small for the initial transition.
If $B$ remains small until the vacuum approaches our vacuum, the situation is one of rapid cascading with each intermediate bubble carried away by the higher vacuum energy background.
Thus the inflating multiverse may decay explosively everywhere to our current low-lying metastable positive vacuum energy universe.

\item{\bf Since such a phase transition does not create particles only a transition to our $\rho_v$ near the peak of the pre-existing matter distribution ($r=0$) will lead to galaxies, planets, and people.}
Bubbles nucleated at $r$ significantly different from zero will have insufficient matter to generate life.  It is possible that only one or only very few bubbles contain intelligent life.  

Even at $r=0$ the matter distribution falls off as $a(t)^{-3}$ so there is, perhaps, an anthropic understanding of an early end to inflation. Neighboring bubbles have very close to the same vacuum energy so there is little effect from bubble collisions contrary to the prediction of the chaotic inflation picture \cite{Johnson}.

Our assumption is contrary to a common untested presupposition (often referred to as the cosmological principle) that our universe does not occupy a privileged position in the multiverse.  

The root mean square radius of the matter distribution from eq.\,\ref{rhom} is
\be
      \sqrt{<r^2>} = \frac{3}{2}\,a(t)\,R\ \quad .
\ee

\item{\bf We are protected from being enveloped in a black hole like region by a future phase transition to zero vacuum energy}
As each bubble grows the energy enclosed approaches the value at which light is trapped.
If there were no energy outside, this critical mass would be the mass of a black hole of that radius and the metric outside would be that of Schwarzschild.  
In the standard cosmology every trajectory through the multiverse, except for a set of measure zero, ends on a black hole.
This and other problematic features discussed in \cite{Banks} derive primarily from the asymptotic features of eternal inflation.
{Banks has shown that transitions between symmetric (constant curvature) spaces of differing vacuum energy do not occur contrary to the usual multiverse picture.  This leaves open the possibility that inhomogeneous models such as ours could support vacuum transitions although bubble nucleation in regions of negligible matter density could be suppressed.}
 
The finite age of the multiverse allows that the limiting radius has not yet been reached.
\be
     \int_{0}^{r} 4\pi {r^\prime}^2 dr^\prime (\rho_m(r^\prime,t)+\rho_v)  + 4 \pi r^2 S /c^2 < \frac{r c^2}{2 G_N} \quad .
\label{BHlimit}
\ee  
Here $S$ is the energy per unit area on the bubble surface.  If it is as small as the estimate of eq.\,\ref{estimates} the surface term can be neglected.  The time at which the visible universe saturates this inequality is given by putting $r=ct_{max}$.  The effect of pressure may be important but has not yet been analyzed.
The condition that this limit has not yet been reached puts a tight limit on the current matter density of the universe.
Whether reaching the limit is incompatible with the continuation of life in the universe is not clear.
\be
     \frac{4 \pi}{3} \rho_c t_{max}^3 = t_{max} /(2 G_N)
\ee
or
\be 
      t_{max} = \sqrt{\frac{3}{8 \pi \rho_c G_N}} = \frac{1}{H_0} = (13.58\,\pm 0.27\,) \displaystyle{Gyr} \quad .
\label{tmax}
\ee
Here $H_0$ is the current value of Hubble's constant which is determined observationally by the ratio of a galaxy's recession speed to its distance from us.
The critical density, $\rho_c$, is determined by the central equation of eq.\,\ref{tmax}.  The numerical value for $H_0$ given in eq.\,\ref{tmax} is from the latest WMAP compilation including other relevant data.
$t_{max}$ should be compared to the current age of the universe, $t_0$.
\be 
    t_0 = \frac{1}{H_0} F(\Omega_m(0), \Omega_\Lambda(0), \Omega_\gamma(0), ..) \quad .
\label{t0}
\ee
The correction factor, $F$, which depends on the current values of various densities is obtained \cite{Jarosik} by integrating back from the current time $t=0$ to the point at which the scale factor goes to zero.  Since the scale factor is only determined up to a constant multiplicative factor we may take $a(0)=1$.  The current values of the Hubble constant and the matter density of the universe
taken from the 2011 Particle Data Group compilation \cite{PDG} are

\be
     h&=&0.702 \pm 0.014\\
     {H_0}^{-1} &=& \frac{9.777752 \displaystyle{Gyr}}{h} = 13.58 \pm 0.27 \,{\displaystyle{Gyr}}\\
     \Omega_m(0) &=& 0.26 \pm 0.02 \quad .
\ee
For $r$ very small and very large we have approximately a De Sitter space so we can use, as an approximation, the usual equations for the scale factor.
Since 
\be
     \frac{\dot{a}}{a} = H(t)
\ee
and, with a sum over species, 
\be
     \frac{\ddot{a}}{a} &=& - \frac{4 \pi G_N}{3} \sum{(\rho_i + 3 p_i)}
       = {H_0}^2 \left(\frac{-\rho_m}{2 \rho_c}+ \frac{\rho_v}{\rho_c}\right)
\ee
we have
\be
     \frac{dH}{dt} = -\frac{3}{2}{H_0}^2\,\Omega_m(t)\quad .
\ee
We may write
\be
      \frac{da}{a H(t)} = dt \quad .
\ee
The age of the universe, $t_0$, is therefore
\be
    t_0 = \int_{0}^{1} \frac{da}{a\,H} 
\ee
where we have written $H$ parametrically as a function of $a$.  Equivalently, $t_0$ can also be defined by
\be
    \int_{-t_0}^{0} dt H(t) = \int_{0}^{1} da/a 
\ee
where $H(-t_0)=\infty$.
If we are not already living in a light trapping region $t_0$ must be less than the $t_{max}$ of eq.\,\ref{tmax}, i.e. the correction factor $F$ in eq.\,\ref{t0} must be less than unity.  The time from now at which our universe becomes light trapping is
\be
   \Delta t = \frac{1}{H_0} (1 - F(\Omega_m(0), \Omega_\Lambda(0), \Omega_\gamma(0), ..)) \quad .
\ee
Assuming that $\Omega_\Lambda(0) = 1 - \Omega_m(0)$ and that the photon and neutrino contributions to the energy content of the universe can be neglected, $\Delta t$ is plotted against $\Omega_m(0)$ in fig.\,\ref{Deltat}. To be consistent with the experimental value, $\Omega_m(0)$ must also be less than $0.28$ and the future lifetime of the universe in the present phase must be less than $0.23$ Gyr.  
{This estimate is two orders of magnitude lower than another estimate \cite{Page} which suggests that the future lifetime against vacuum decay could be of order $20$ Gyr.}

\item{\bf Eventually, a bubble of zero vacuum energy should form within our universe and expand to convert the entire universe to this presumably supersymmetric state.}  At this point the interior  of the bubble may never reach the light trapping limit since the first term in eq.\,\ref{BHlimit} is limited by $M$ and the vacuum energy term drops to zero. 
The surface term can never contribute to light trapping since it is itself moving out at the speed of light.  
Other regions devoid of significant matter densities will undergo the transition to zero vacuum energy without producing matter. 
If we are not destined to be enveloped in a light trapping region, the current model predicts
a cosmologically proximate end to the current era in a transition to a universe of zero vacuum energy.

\item{\bf The zero vacuum energy state is the ground state of the universe.} If there are local minima of negative vacuum energy with the same number of degrees of freedom as our universe, there might eventually be a transition to one or more of these leading to a big crunch.  However, Banks \cite{Banks}
has given strong arguments that this will not happen.
In the standard picture where the mass is not finite, the eventual formation of a light trapping region  is then unavoidable.
Assumption 9 and the previous one are motivated by avoiding the light trapping limit although it is not clear whether such a limit would be inimical to life or pose other problems to the theory. 

\end{enumerate} 

\begin{figure}[h]
\begin{center}
\epsfxsize= 3in 
\leavevmode
\epsfbox{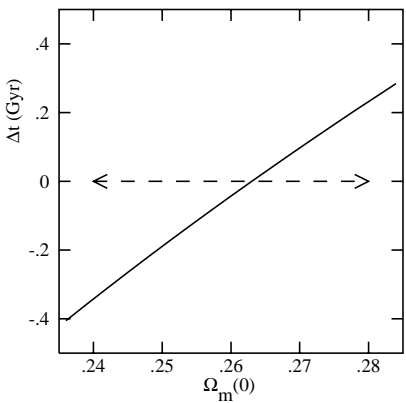}
\end{center}
\caption{\baselineskip=10pt The time from now before our universe becomes light trapping is plotted against the current matter density, $\Omega_m(0)$.  The experimental range of this quantity is ‌shown by the dashed line.  As indicated, if the universe is not already light trapping, ($\Delta t>0$), $\Omega_m(0)$ must be greater than $0.2645$.  }  
\label{Deltat}
\end{figure}

\section{Conclusion}

In the current model we have introduced several free parameters for which we have only given preliminary estimates.  
The model can be falsified if inconsistencies develop and could be confirmed if systematic matter inhomogeneities are found within the Hubble length.  The thesis of this paper is that
what is called the cosmological principle may be an over-generalization of observations at lower
distance scales.  Our proposal is anti-Copernican in the sense that our inhabited bubble universe is at the center of the multiverse.  
As $R\,\rightarrow\,\infty$, the model approaches the standard landscape cosmology.  However, for any finite $R$, the model has the following advantages over the standard picture.

\begin{itemize}
\item{we can avoid an infinite number of human, quasi-human, and monster populations in the string landscape and we have, at least,
a reduced production of Boltzmann Brains and other bizarres entities,}
\item{with a finite amount of matter in the multiverse we have no measure problem,}
\item{we can unify an inhomogeneous matter distribution with an apparent FRW metric,} 
\item{a simple solution to the baryon asymmetry problem is possible,}
\item{a clear definition of inertial frames is obtained,}
\item{the earth rest frame is approximately inertial,} 
\item{we can understand the current order of magnitude equality between matter and dark energy densities,} and 
\item{we can at least potentially avoid becoming enveloped by a black hole like state.} 
\end{itemize}

At present it may be considered a matter of individual taste as to whether these advantages outweigh the required paradigm shift.  In particular, many cosmologists seem at peace with the prospect that we and the entire human history of life on earth may be playing out an infinite number of times in other parts of the multiverse.  At a minimum the present paper could be considered a challenge to cosmologists to prove the uniqueness of the infinite cloning prediction or to find other alternative scenarios which avoid the infinite cloning and preserve some or all of the conceptional advantages of our model while also preserving the string landscape explanation for the smallness of the vacuum energy. 
 
It is left to future work to consider the general relativistic corrections to the picture presented here and to incorporate the effects of non-zero pressure. In the process, modifications to the matter density of eq.\,\ref{rhom} that preserve the properties of eqs.\,\ref{delta} and \ref{Mtot} may become evident.

{\bf Acknowledgements} We acknowledge useful discussions of the matter presented here with Larry Ford at Tufts University.  The research of LC was supported in part by the DOE under grant DE-FG02-10ER41714 and that of GG under DOE grant DE-FG02-92ER40702.

\end{document}